\documentstyle[12pt]{article}
\begin{document}
\begin{titlepage}
October 1996 \hfill\hfill UMH-MG-96/01
\vskip 3.cm
\begin{center}
{\LARGE\bf Massive spin 2 propagators on \\
de Sitter space}
\vskip 27.mm 
{Cl. Gabriel\footnote{Aspirant du FNRS} \footnote{E-mail gabriel@sun1.umh.ac.be}, 
Ph. Spindel\footnote{E-mail spindel@sun1.umh.ac.be}}\\
\vskip 1 cm
 {\em M\'ecanique et Gravitation}\\
 {\em Universit\'e de Mons-Hainaut}, \\
{\em 15, avenue Maistriau, B-7000 Mons, Belgium}
\end{center}
\vskip 1 cm
\begin{center}
\begin{abstract}

We compute the Pauli-Jordan, Hadamard and Feynman propagators for the massive 
metrical perturbations on de Sitter space. They are expressed both in terms
of mode sums and in invariant forms. 
\end{abstract}
\end{center}

\end{titlepage}
\pagebreak
\baselineskip=24pt
\renewcommand\thesection{\Roman{section}}
\section{Introduction}
\par The apparently simplest model of space-time beyond Minkowski space is de Sitter
space. Its high (maximal) degree of symmetry makes it the typical framework 
to investigate quantum field theory outside flat space.\\
\noindent One of the essential 
ingredients of quantum field theory are the various Green functions of the fields. For the scalar field on de Sitter space, the first work
on the subject is, in our knowledge, the paper of G\'eh\'eniau and Schomblond
\cite{GeSc}. These authors have used the harmonicity property of de Sitter space, 
i.e. the possibility to solve Klein-Gordon equation by a function depending only on the 
geodesic distance, to obtain the expression of the Pauli-Jordan propagator $\Delta(x,y)$.
Soon after, Cahen, G\'eh\'eniau, G\"unther and Schomblond \cite{CaGGSc} have obtained
the expression of the analogous Green function $S(x,y)$ for the spinorial field. Their method
consisted essentially to compute $S(x,y)$ first with $y$ fixed at the origin of
a coordinate patch, and then, to generalize the expression so obtained for an arbitrary
couple of points by using parallel transport. Later, Schomblond and one of 
the authors of this work (Ph.S) have obtained a Fock space description
of the scalar \cite{ScSp1} , spinorial and vectorial \cite{ScSp2} propagators
by computing them as mode sums. The main result in \cite{ScSp1} was that by
imposing invariance conditions (with respect to the isometries of the space) and fixing 
the behaviour at short distances of the propagator, a uniqueness theorem holds. All ambiguities
about the definition of particles, expressed by arbitrary Bogoljubov's
transformations, are resolved. At the same time, Candelas and Raine \cite{CaRa}
obtained a similar result using the harmonicity properties of the space to 
write a Schwinger representation of the Feynman propagator depending only on
the geodesic distance and satisfying a regularity boundary condition (imposed 
on the kernel of the Schwinger representation).\\
\noindent Let us emphasize that, for the massless scalar field ($\Box \Phi=0$),
it is impossible to obtain a fully de Sitter invariant vacuum state. This result
was first noticed by Spindel \cite{SpTh} and independently rediscovered in 1982
by Vilenkin and Ford \cite{ViFo}. It has been discussed in details by Allen and
Folacci \cite{Allen1,AlFo}. Allen has also obtained explicitly de Sitter invariant representations for the Green functions of the spinorial, vectorial and gravitational
fields \cite{Allen2,Allen3,AlTu}, under the assumption a priori that these Green functions could be expressed in terms of
products of functions of the geodesic distance with maximally symmetric bispinors or bitensors.\\ 
\noindent A drawback of this geometrical construction of the Green functions is 
that we have no information about the existence of an underlying Fock space, i.e. a vacuum state 
such that expectations values of fields products with respect to it give the corresponding 
Green functions. Actually, for the gravitational field, the situation is the same as for the massless 
spin 0 field \cite{GeSc}: there exists de Sitter invariant Green functions, but no corresponding
vacuum state. Representations of these Green functions have been obtained by
Allen and Turyn \cite{AlTu,Allen4} and by Antoniadis and Mottola \cite{AnMo}. Both
results, which differ only by gauge choice, are obtained as analytic continuation 
of Green functions built on the Euclidean 4-sphere. A direct evaluation of the
gravitational propagator as mode sums in physical space (de Sitter space) has been
done by Tsamis and Woodard \cite{TsaWoo}. Their construction leads to a result
analogous to the one already obtained for the massless, minimally coupled scalar
field: there is no de Sitter invariant vacuum state for the massless spin 2 field.\\
\noindent In this note, we shall perform the calculation of the propagator
for the ``massive spin 2 field". More precisely, we consider the gravitational perturbation
field equations introduced by Lichnerowicz \cite{lichne}, which corresponds to
a mixture of spin 0 and spin 2 fields. In section 2, we summarize the field
equations and remind the mode sums representations of the various propagators.
In the third section, we specialize the field equations on (3+1) de Sitter space
and solve them explicitly on a half de Sitter space. We then obtain the propagators by summing modes.
In section 4, we establish coordinate-free and manifestly $O(4,1)$ invariant
representations of the propagator. In section 5,we discuss quickly the analytic continuation
of the modes and propagators on the full de Sitter space.\\ 
\section{Metric perturbation on Einstein space}
Following Lichnerowicz \cite{lichne}, we write the equations of motion for  ``massive
metric perturbations" on an Einstein background space 
$(R_{\alpha \beta} = \Lambda
g_{\alpha \beta})$ as
\begin{eqnarray}
\delta R_{\alpha \beta} (\mbox{\bf h}) & = & \mu h_{\alpha \beta} \label{1}
\end{eqnarray}
where $\delta R_{\alpha \beta}$ ({\bf h}) denotes the terms linear 
in the components of the tensor {\bf h} 
in the expansion of the Ricci tensor evaluated on
the metric {\bf g}' = {\bf g} + {\bf h}. 
The factor $\mu$ on
the right hand side of eq.(\ref{1}) is related to a mass term as 
$M^2 =2 (\Lambda - \mu)$. This definition of mass is meaningful because
$M^2=0$ corresponds to pure gravity. Hereafter, we shall denote by ${m \over R^2}=-2\mu$ the
eigenvalues of the quadratic Casimir operator $I_1$ considered by B\"orner and
D\"urr \cite{BoDu} who have used the notations $m^2=-2\mu$ and $m_0=i\nu$ (see eq.(\ref{46})).
\par\noindent From the Bianchi identities, written for the metric 
{\bf g}', we
deduce that solutions of eq.(\ref{1}) satisfy automatically 
the de Donder conditions:
\begin{eqnarray}
&& (\mu - \Lambda) \nabla_{\alpha} \hbar^{\alpha \beta}  =  0 \;\quad . \label{2}
\end{eqnarray}
In these equations, as in the rest of this paper, we have introduced the Einsteinian conjugate tensor $\hbar_{\alpha\beta} 
= h_{\alpha \beta} - \frac 12 g_{\alpha \beta} h_\mu^\mu$, and the
covariant derivatives refer to the Levi-Civitta connection 
built on the metric
$\bf g$ which is used to lower and raise the indices. 
Consequently, eq.(\ref{1}) is
equivalent to the system constituted by eq.(\ref{2}) and
\begin{eqnarray}
\Box h_{\alpha \beta} + 2 R_{\alpha \sigma \beta \rho} 
h^{\sigma \rho} - 2 (\Lambda - \mu) h_{\alpha \beta} = 0 \;\quad . \label {3}
\end{eqnarray}
A particular solution of eqs (\ref{2},{3}) is given by:
\begin{eqnarray}
h_{\alpha \beta}= \nabla_\alpha \nabla_\beta \phi + (\Lambda - \mu) g_{\alpha
\beta} \phi \label{4}
\end{eqnarray}
when $ \phi$ satisfies the scalar field equation~:
\begin{eqnarray}
\Box \phi + 2 \mu \phi = 0 \quad.\label {5}
\end{eqnarray}
This scalar field is proportional to the trace:
\begin{eqnarray}
h_\mu^\mu = ( 4 \Lambda - 6 \mu) \phi \qquad , \label{6}
\end{eqnarray}
and describes the spin 0 content of $\bf h$, which is a mixture of spins 0 and 2.\\ 
\pagebreak
\noindent Equation (\ref{1}) can be derived from the Lagrangian :
\begin{eqnarray}
{\cal {L}} & = & {1 \over 2} \nabla^{\alpha} h^{\beta \gamma} \nabla^\mu h^{\nu \rho}
\left( g_{\alpha \mu} g_{\beta \nu} g_{\gamma \rho} - g_{\alpha \rho}  g_{\beta \mu}
 g _{\gamma \nu} - g_{\mu \beta} g_{\nu \alpha}
g_{\rho \gamma} \right.\nonumber \\
&& \left.+ g_{\mu \nu} g_{\rho \alpha} g_{\beta \gamma} + g_{\alpha
\beta} g_{\gamma \mu} g_{\nu \rho} - g_{\alpha \mu} g_{\beta \gamma} g_{\nu \rho}
\right) - {\mu \over 2}  h^{\alpha \beta} h^{\mu \nu} \left( g_{\alpha \mu}
g_{\beta \nu} \right.\nonumber \\
&& \left.+ g_{\alpha \nu} g_{\beta \mu} - g_{\alpha \beta} g_{\mu
\nu}\right) \equiv {1 \over 2}\nabla^\alpha h^{\beta \gamma} \nabla^\mu h^{\nu
\rho} \; Q_{\alpha \beta \gamma, \mu \nu \rho} - {\mu \over 2} h^{\alpha
\beta} h^{\mu \nu} P_{\alpha \beta, \mu \nu} \label{7}  
\end{eqnarray}
which is unique, up to trivial transformations (rescaling and addition of
divergences). From it we deduce the expression of a conserved current, a
sesquilinear form on the space of complex solutions of eq.(\ref{1}):
\begin{eqnarray}
 J_\alpha (h, k) & = & i (h^{* \beta \gamma} \nabla^\mu k^{\nu \rho} - k^{\beta
\gamma} \nabla^\mu h^{* \nu \rho} ) Q_{\alpha \beta \gamma, \mu \nu \rho} \label{8} 
\end{eqnarray}
and, by integration on a Cauchy surface $\Sigma$, a symplectic
structure: 
\begin{eqnarray}
h* k & = & \int_{\Sigma} d \sigma^\alpha J_\alpha (h, k) = - (k* h)^* = -(k^* * h^*)
\label {9}.
\end{eqnarray}
If $ \{^Ah\}$ is a complete set of positive frequency modes, labelled by the
index $A$, and satisfying the relations
\begin{eqnarray}
\,^A h* ^Bh &=& \delta_{AB} \nonumber \\
\,^A h^* * ^Bh^* &=& - \delta_{AB} \nonumber \\
\,^A h^* * ^B h &=& 0 =\; ^A h * \; ^B h^* \label{10}
\end{eqnarray}
we may obtain as mode sums the usual Green functions of the quantum field $\bf{\hat h}$ associated to the classical field $\bf {h}$. The Pauli-Jordan propagator $ \Delta(x, y)$, defined
by the commutator $ [ {\bf {\hat h}} (x), {\bf {\hat h}} (y) ] $, is given by:
\begin{eqnarray}
\Delta (x,y) &=& - i \sum_A \;^Ah (x) ^Ah^* (y) - ^Ah^* (x) ^Ah (y) \quad ,
\end{eqnarray}
while the symmetric (often called Hadamard) propagator $ \Delta^1 (x,y)$, defined as the vacuum expectation value of the anti-commutator $
\langle \{ {\bf {\hat h}} (x),{\bf {\hat h}} (y) \} \rangle$, is:
\begin{eqnarray}
\Delta^1 (x,y) &=& \sum_A \;^Ah (x) ^Ah^* (y) + ^Ah^* (x) ^Ah (y) \quad ,
\end{eqnarray}
and the Feynman propagator $ \Delta_F (x, y)$:
\begin{eqnarray}
\Delta_F (x, y) &=& {1 \over 2} [ \Delta^1 (x,y) + i \epsilon (x, y) \Delta (x,
y) ] \; , \label{11}
\end{eqnarray}
\noindent where $ \epsilon (x, y) = \pm 1$ according as the point $x$ is in
future or the past of $y$.

\section{Field equations on de Sitter space}
The four dimensional de Sitter space $ H^4$ can be seen as the
homogeneous coset space  \mbox{O(4,1)/O(3,1)}, i.e as the sphere 
of equation:
\begin{eqnarray}
\eta_{AB} X^A X^B = R^2 \qquad (A, B= 0, ..., 4) \label{12}
\end{eqnarray}

\bigskip

\noindent imbedded in a five dimensional (flat) Minkowski space $ M^5$. Using the
parametrization:

\begin{eqnarray}
\lambda = {R^2 \over X^4 - X^0} \qquad ,\qquad x^i = {R X^i\over X^4 - X^0} \quad 
(i = 1, 2, 3) \label{13}
\end{eqnarray}

\bigskip

\noindent the metric induced on $H^4$ reads as

\begin{eqnarray} 
g = {R^2 \over \lambda^2} \left(- d \lambda^2 + \sum_i
(dx^i)^2 \right) \; . \label{14}
\end{eqnarray}

\bigskip

\noindent De Sitter space-time is a space of constant curvature:
$R_{\alpha \beta \gamma \delta} = {\Lambda \over 3} (g_{\alpha \gamma} g_{\beta \delta} -
g_{\alpha \delta} g_{\beta \gamma})$, 
the relation between the cosmological constant $ \Lambda$ and the
radius $R$ of the space being: $\Lambda = 3 / R^2 \;$.
\bigskip

\noindent On de Sitter space eq.(\ref{3}) becomes equivalent to

\begin{eqnarray}
\Box \;\hbar_{\alpha \beta} + \left({ 8 \over 3} \Lambda - 2 \mu \right)
\hbar_{\alpha \beta} - {2 \over 3} \Lambda g_{\alpha \beta} \hbar_\tau^\tau = 0
\; . \label {17}
\end{eqnarray}

\bigskip
\noindent Hereafter we shall in a first step restrict ourselves to the chart
$\lambda > 0$, which covers only one half of the full de Sitter space. 
It is the domain corresponding to the causal past of a physical observer 
(region ${\cal{O}}$ on fig.1).\\

\noindent To solve the system of equations (\ref {17}) we have found useful to use
the rescaled quantities introduced in \cite{TsaWoo}:

\begin{eqnarray}
k_{\mu \nu} = {\lambda^2 \over R^2} \; \hbar_{\mu \nu} \label {18}
\end{eqnarray}

\bigskip

\noindent and to pass to the Fourier transformed variables  :

\begin{eqnarray}
K_{\mu \nu} (\lambda, \vec p) = {1 \over (2 \pi)^{3/2}} \int d^3 p \; e^{-i
\vec p. \vec x} k_{\mu \nu} (\lambda, \vec x) \; . \label{19}
\end{eqnarray}

\bigskip
\noindent where $ \vec p. \vec x = \sum_i p^i x^i $.
\bigskip

\noindent Moreover it is natural to express these Fourier components in a local frame
adapted to the vector $ \vec p .$ To this aim, we introduce the four vectors $
u_{(\alpha)} ( \vec p)$ whose components with respect to the natural coordinate 
frame $(\partial_\lambda,\partial_{x^i})$ are:

\begin{eqnarray}
u_{(0)}^\alpha &=& ( 1, 0, 0, 0) \nonumber \\
u_{(3)}^\alpha &=& (0, p^i / p) \qquad, \quad p = \sqrt{\sum_i (p^i)^2}  \label{20} \\
u_{(1, 2)}^\alpha &=& (0 ,\varepsilon_{(1, 2)}^i ) ,\quad \mbox{\rm with}\quad \sum_i p^i \varepsilon^i_{(1, 2)}
= 0 \; .\nonumber  
\end{eqnarray}

\bigskip

\noindent They  satisfy the orthogonality relations

\begin{eqnarray}
g_{\mu \nu} u_{(\alpha)} ^\mu ( \vec p) u_{(\beta)} ^\nu (\vec p) = {R^2
\over \lambda^2} \eta_{\alpha \beta} \quad . \label{21} 
\end{eqnarray}

\bigskip

\noindent We shall also make use of the projector on the spacelike directions
orthogonal to $ \vec p$:

\begin{eqnarray}
\perp_\mu ^\nu &=& \delta_\mu ^\nu + {\lambda^2 \over R^2} \left( u_{(0) \mu}
u_{(0)}^\nu - u_{(3) \mu} u_{(3)} ^\nu \right) \nonumber \\
&=& {\lambda^2 \over R^2} \left( u_{(1) \mu} u_{(1)}^\nu + u_{(2)\mu }u_{(2)}^\nu
\right) \quad . \label{22}
\end{eqnarray}

\bigskip

\noindent Once expressed in this frame, the components of (\ref{19}) split into
longitudinal and transverse parts :

\begin{eqnarray}
K^L &=& u_{(3)} ^\alpha K_{0 \alpha} = {p^i \over p} K_{0i} \qquad , \\
K_\alpha^\perp &=& (0, K_i ^\perp) = (0, \perp_i ^j K_{0j}) \qquad , \\
K^{LL}&=& u_{(3)} ^\alpha u_{(3)} ^\beta K_{\alpha \beta} \qquad ,\\
K_\alpha ^{L\perp} &=& (0, K_i^{L\perp} ) = u_{(3)} ^\mu \perp_\alpha^\nu K_{\mu \nu}\qquad ,\\
K_{\alpha \beta} ^{\perp \perp} &=& \perp_\alpha ^\mu \perp_\beta^\nu K_{\mu \nu},\\
K_{0 \alpha} ^{\perp \perp} &=& K_{\alpha 0}^{\perp \perp} = 0 \qquad .
\label{27} 
\end{eqnarray}

\bigskip

\noindent Conversely, the components of (\ref{19}) with respect to the natural frame read

\begin{eqnarray}
K_{0i} &=& u_{(3)} ^i K^L + K_i ^\perp \qquad ,\\
K_{ij} &=& u_{(3)} ^i u_{(3)} ^j K^{LL} + u_{(3)} ^i K_j^{L\perp} + u_{(3)} ^j
K_i^{L\perp} + K_{ij} ^{\perp \perp}. \label{29}
\end{eqnarray}

\bigskip

\noindent In terms of these variables eq.(\ref {2}) splits into:

\begin{eqnarray}
\dot K_{00} - i\; p \; K^L - {4 \over \lambda} K_{00} - {1 \over \lambda}K & = & 0 \label{30} \\
\dot K^L - i\; p \; K^{LL} - {4 \over \lambda} K^L & = & 0  \label{31} \\
\dot K_{i} ^\perp - i \; p\;  K_i^{L \perp} - {4 \over \lambda} K_{i}^\perp &=& 0 \label{32}
\end{eqnarray}

\bigskip

\noindent where we have denoted by dots derivatives with respect to $ \lambda$
and by $K$ the trace of $ \hbar_{\mu \nu}$ :

\begin{eqnarray}
K = - K_{00} + \sum_i K_{ii} = \hbar_\mu ^\mu = - h_\mu^\mu \quad . \label {33}
\end{eqnarray}

\bigskip

\noindent Eqs (\ref {17}) written with $m = - 2 \mu R^2$ become:

\begin{eqnarray}
\ddot{K}_{00} - {6 \over \lambda} \; \dot K_{00} + 
\left(p^2 + {16 + m \over \lambda^2}
 \right) K_{00} + {4 \over \lambda^2} K &=& 0 \; \label{34} \\
\ddot{K}^L - {4 \over \lambda} \dot K ^L - {2i \over
\lambda} p K_{00} + \left( p^2 + {10 + m \over \lambda^2}\right) 
K^L &=& 0 \;\label{35} \\
\ddot{K}_j^\perp - {4 \over \lambda} K_j^\perp + \left( p^2 + {10 + m \over
\lambda^2} \right) K_j^\perp &=& 0 \;\label{36} \\
\ddot{K}^{LL} - {2 \over \lambda} \dot K^{LL} + \left( p^2 + {6 + m \over
\lambda^2} \right) K^{LL} - {4i \over \lambda} p K^L  - {2 \over \lambda^2} (K
+ K_{00}) &=& 0 \; \label{37} \\
\ddot{K}_j^{L\perp} - {2 \over \lambda} \dot K_j^{LL} + \left(
p^2 + {6 + m \over \lambda^2}\right) K_j ^{L\perp} - {2i \over \lambda} p
K_j^\perp &=& 0 \; \label{38} \\
\ddot{K}_{jl}^{\perp \perp} - {2 \over \lambda} \dot K_{jl}^{\perp
\perp} + \left( p^2 + {6 + m \over \lambda^2}\right) K_{jl}^{\perp \perp} - {2
\over \lambda^2} \perp_{jl} (K_{00} + K) &=& 0 \; \label{39}
\end{eqnarray}

\bigskip

\noindent Summing the appropriate equations (\ref{34}, \ref{37}, \ref{39}) we recover
with the help of eqs (\ref {30}, \ref{31}, \ref{32}) the trace equation (\ref{5}) written in 
terms of its Fourier transformed variable:

\begin{eqnarray}
\ddot{K} - {2 \over \lambda} \dot K + \left( p^2 + {m \over \lambda^2}
\right) K = 0 \quad . \label{40} 
\end{eqnarray}

\bigskip

\noindent The general solution of this equation is given by a combination of
Hankel functions (see \cite{Abra}):

\begin{eqnarray}
K ( \lambda, \vec p) = (\lambda p)^{3/2} \left[ a (\vec p) {\cal {H}}_{\nu_0}
^{(1)} ( \lambda p) + b ({\vec p}) {\cal H}_{\nu_0} ^{(2)} ( \lambda p) \right]
\label{41} 
\end{eqnarray}

\bigskip

\noindent with

\begin{eqnarray}
\nu_0 = i \; \sqrt{m - {9 \over 4}} \label {42} 
\end{eqnarray}

\bigskip

\noindent and

\begin{eqnarray}
{\cal H}_\nu^{(1)} ( \lambda p) = e^{i \nu{\pi \over 2}} \; H _\nu^{(1)} (\lambda p)
= [ {\cal H}_\nu^{(2)} ( \lambda p) ]^* \; , \label {43}
\end{eqnarray}

\bigskip

\noindent where $ H_\nu^{(1)} , H_\nu^{(2)}$ are the usual Hankel functions  defined in
\cite{Abra}.
\bigskip

\noindent From eqs (\ref {4}, \ref{5}) and eq.(\ref {34}) we obtain (assuming $m + 4 \neq
-n(n+1), \quad n \in {\bf Z} $, see comment after eq.(\ref{plus}) ):

\begin{eqnarray}
K_{00} = Q - {\lambda^2 \over 3 (m + 4)} \; \left( {3 \over \lambda} \dot K
- (p^2 - {3 \over \lambda^2}) K \right) \label {44} 
\end{eqnarray}

\bigskip

\noindent where $Q$ is the general solution of the homogeneous part $ (K = 0)$ of
eq.(\ref {34}):

\begin{eqnarray}
Q = (\lambda p)^{7/2} \left[ c (\vec p) {\cal H}_\nu^{(1)} ( \lambda p) + d
(\vec p) \; {\cal H} _\nu ^{(2)} ( \lambda p) \right] \label {45}
\end{eqnarray}

\bigskip

\noindent and here

\begin{eqnarray}
\nu = i \;\sqrt{m + {15 \over 4}} \quad .\label {46} 
\end{eqnarray}

\bigskip

\noindent From these solutions and eq.(\ref {30}) we obtain algebraically the $ K^L$
component:

\begin{eqnarray}
K^L = {-i \over p} \left [ \dot Q - {4 \over \lambda} Q + {\lambda^2 p^2
\over 3 (m + 4)} \left( \dot K + {1 \over \lambda} K \right) \right] \; . \label{47} 
\end{eqnarray}

\bigskip

\noindent In the same way eq.(\ref {31}) gives immediately

\begin{eqnarray}
K^{LL} = {1 \over p^2} \left[ {2 \over \lambda} \dot Q + (p^2 + {m - 4 \over
\lambda^2}) Q \right] - {\lambda^2 \over 3 ( m + 4)} \left[ {1 \over \lambda}
\dot K - (p^2 + {m + 3 \over \lambda^2} ) K \right] \; . \label {48}
\end{eqnarray}

\bigskip

\noindent The equation (\ref {36}) is decoupled from the others. Its general solution is :

\begin{eqnarray}
K_j ^\perp = ( \lambda p)^{5/2} \left[ c_j ( \vec p) {\cal H}_\nu^{(1)} (
\lambda p) + d_j (\vec p) {\cal H} _\nu ^{(2)} ( \lambda p) \right] \; .\label
{49} 
\end{eqnarray}
\bigskip

\noindent where $ c_j ( \vec p)$ and $ d_j ( \vec p)$ are 3-vectors orthogonal to
$\vec p $ and $\nu$ is again given by eq.(\ref{46}). 
This leads directly, thanks to eq.(\ref {32}), to

\begin{eqnarray}
K_j ^{L \perp} = - {i \over p} \left( \dot K_j ^\perp - {4 \over \lambda}
K_j ^\perp \right) \; . \label {50} 
\end{eqnarray}

\bigskip

\noindent Finally it remains to solve eq.(\ref{39}) for $ K_{j l} ^{\perp \perp}$. The
general solution of the homogenous part is still given by a combination of Hankel
functions of the same index $\nu$ (see eq.(\ref{46})):

\begin{eqnarray}
Q_{j l} ^{\perp \perp} = ( \lambda p)^{3/2} \left[ c_{jl} ( \vec p) \; {\cal
H}_{\nu}^{(1)} (\lambda p) + d_{jl}  (\vec p) \; {\cal H} _{\nu}^{(2)} (\lambda p)
\right] \; , \label {51} 
\end{eqnarray}

\bigskip

\noindent while a particular solution can be expressed in terms of $Q$ and $K$,
leading to the general solution:

\begin{eqnarray}
K_{jl} ^{\perp \perp} = Q_{jl} ^{\perp \perp} - \perp_{jl} \left[ {1 \over 3
(m + 4)} \left( \lambda \dot K - (3 + m)  K \right) \right. + \left.{1 \over 2
p^2 \lambda^2} ( 2 \lambda \dot Q + (m - 4) Q) \right] \label{52}
\end{eqnarray}

\bigskip
\noindent The integration constants $c_{jl}$ (and similarly $d_{jl})$ can be
expressed with the help of the projector (\ref{22}) as:

\begin{eqnarray}
c_{jl} ( \vec p) = \sum_{m, n} ( \perp_{jm} \perp_{ln} - {1 \over
2} \perp_{jl} \perp_{mn}) {\cal E}_{mn} ( \vec p) \label {53}
\end{eqnarray}

\bigskip

\noindent where $ {\cal E}_{mn} ( \vec p)$ is arbitrary. This form insures that
$ c_{jl}$ is transverse to $ \vec p$ and traceless in order to verify eq.(\ref{29}).
\bigskip

\noindent To be complete we have still to check that the longitudinal components
(\ref {47}, \ref{48}, \ref{50}), obtained from the divergence equations, are really solutions of
the second order equations (\ref {35}, \ref{37} and \ref{38}). They are!

\noindent Now we may write a complete set of modes, 
solutions of eqs (\ref{2},\ref{17}) as:

\begin{eqnarray}
\hbar_{\mu \nu} ( \lambda, \vec x, \vec p) = {e^{i \vec p. \vec x} \over (2
\pi)^{3/2}} \; {R^2 \over \lambda^2} \; K_{\mu \nu} (\lambda, \vec p) \;
. \label{54}
\end{eqnarray}

\bigskip  
\noindent However instead of considering their natural  components it is
more useful to split the modes according to their spin contents. Indeed, such a
decomposition leads automatically to orthogonal modes and it will just remain
to normalize them. So we shall write the general complex solution
of the field equations as

\begin{eqnarray}
\hat {\hbar}_{\mu \nu} ( \lambda, \vec x) = \sum_I \int d^3 p [ a_I ( \vec p)
\hbar_{\mu \nu} ^I ( \lambda, \vec x, \vec p)  +  b_I^+ ( \vec p) \hbar_{\mu
\nu} ^{I*} ( \lambda, \vec x, \vec p) ]\; \label{55}
\end{eqnarray}

\bigskip
\noindent where the index $I$ runs over six values corresponding to the spin 0
and spin 2 content of the field (See appendix for the explicit form of the modes).

\bigskip

\noindent If we assume that the amplitudes $ a_I ( \vec p), b_I (\vec p)$ are
operators obeying usual canonical commutation relations:

\begin{eqnarray}
[ a_I ( \vec p), a_{I^{\prime}}^{\dagger} ( \vec p^{\, \prime})] &=& \delta_{II^{\prime}} 
\delta ( \vec p - \vec p^{\, \prime})  \nonumber \\
\left[ b_I ( \vec p), b_{I^{\prime}}^{\dagger} ( \vec p^{\, \prime})\right] &=& \delta_{II^{\prime}} 
\delta ( \vec p - \vec p^{\, \prime}) \nonumber \\ 
\left[ a_I ( \vec p), a_{I^\prime} ( \vec p^{\, \prime})\right
] &=& [ b_I (\vec p), b_{I^{\prime}} ( \vec p^{\, \prime}) ] = [ a_I ( \vec p), b_I ( \vec p^{\, \prime}) ] = ... = 0  \quad, \label{56}
\end{eqnarray} 

\bigskip

\noindent the modes $ \hbar_{\mu \nu} ^I$ have to be normalized to $ \delta^3 (
\vec p - \vec p^\prime)$. 

\noindent Anticipating on the discussion of section V, we impose now that all
modes depends only on $ {\cal H}^{(2)}$ functions, with their various "$d$" coefficients
equal to 1. This choice is compatible with the commutation relations (\ref{56}). It results
from the requirements that $ 1^0$) the Green functions are de Sitter invariant $2^0$)
they have the same short distance behaviour as in flat space.
The coefficients of normalization of the modes with respect to the scalar
product (\ref{9}) are displayed in the appendix. Inserting these modes in the general
expression of the Green functions (\ref{11}) we obtain

\footnotesize
\begin{eqnarray}
\Delta_{00, 0^\prime0^\prime} (x; y) & = & \frac{R^4}{6(m+4)(m+\frac{15}2)}(\frac 3\lambda 
\partial _\lambda -\vec \nabla _x\cdot \vec \nabla _y+\frac 3{\lambda ^2})
(\frac 3{\lambda ^{\prime }}\partial _{\lambda ^{\prime }}-\vec \nabla _x\cdot \vec
 \nabla _y 
 +\frac 3{\lambda ^{2\,\prime }})\Delta _{\nu_0}(p)\nonumber \\
 &&  +
\frac{2R^4}{3(m+6)(m+4)}(\vec \nabla _x\cdot \vec \nabla _y)^2\Delta _{\nu} (p)
\label{57} \\ 
\Delta_{00, 0^\prime i^\prime} ( x; y ) &=& \frac{\,R^4}{6(m+4)(m+\frac{15}2)}
\partial _{i^{\prime }}(\frac 3\lambda \partial _\lambda -\vec \nabla _x\cdot \vec
 \nabla _y+\frac 3{\lambda ^2})(\partial _{\lambda ^{\prime }}+
\frac 1{\lambda ^{\prime }})\Delta _{\nu_0}(p)
\nonumber \\
&& -\frac{2R^4}{3(m+6)(m+4)}
\partial _{i^{\prime }}(\vec \nabla _x\cdot \vec \nabla _y)(\partial 
_{\lambda ^{\prime }}-\frac 2{\lambda ^{\prime }})\Delta _\nu (p) \label{58}\\
\Delta_{00, i^\prime j^\prime} ( x, y) &=& 
\frac{\,R^4}{6(m+4)(m+\frac{15}2)}\eta _{i^{\prime }j^{\prime }}(\frac 3\lambda \partial _\lambda -\vec \nabla _x\cdot
 \vec \nabla _y+\frac 3{\lambda ^2})(\frac{\partial _{\lambda ^{\prime }}}
{\lambda ^{\prime }}-\frac{3+m}{\lambda ^{2\prime }})\Delta _{\nu _0}(p)
\nonumber \\
&& +\frac{\,R^4}
{6(m+4)(m+\frac{15}2)}\partial _{i^{\prime }}\partial _{j^{\prime }}(\frac 3\lambda 
\partial _\lambda -\vec \nabla _x\cdot \vec \nabla _y+\frac 3{\lambda ^2})
\Delta _{\nu _0}(p)\nonumber \\
&& -\frac{2R^4}{3(m+6)(m+4)}\eta _{i^{\prime }j^{\prime }}(\frac{\partial _{\lambda ^{\prime }}}
{\lambda ^{\prime }}+\frac m{2\lambda ^{2\prime }})(\vec \nabla _x\cdot \vec 
\nabla _y)\Delta _\nu (p)\nonumber\\
&& -\frac{2R^4}{3(m+6)(m+4)}\partial _{i^{\prime }}
\partial _{j^{\prime }}(\frac 3{\lambda ^{\prime }}\partial _{\lambda ^{\prime }}
+\vec \nabla _x\cdot \vec \nabla _y+\frac{3m}{2\lambda ^{2^{\prime \,}}})
\Delta _\nu (p)\label{59}\\
\Delta_{0i, 0^\prime j^\prime} ( x, y) &= &\frac{\,R^4}{6(m+4)(m+\frac{15}2)}\partial _i
\partial _{j^{\prime }}(\partial _\lambda +\frac 1\lambda )(\partial _{\lambda ^{\prime }}
+\frac 1{\lambda ^{\prime }})\Delta _{\nu _0}(p)\nonumber\\
&& +\frac{2R^4}{3(m+6)(m+4)}
\partial _i\partial _{j^{\prime }}(\partial _\lambda -\frac 2\lambda )(\partial _{\lambda ^
{\prime }}-\frac 2{\lambda ^{\prime }})\Delta _\nu (p)\nonumber\\
&& +\frac{R^4}{2\,(m+6)}\eta _{i\,j^
{\prime }}(\vec \nabla _x\cdot \vec \nabla _y)\frac 1{\lambda \,\lambda ^
{\prime }}\Delta _\nu (p)\nonumber \\
&&-\frac{R^4}{2\,(m+6)}\partial _i\partial _{j^{\prime }}
\frac 1{\lambda \,\lambda ^{\prime }}\Delta _\nu (p)\label{60}\\
\Delta_{0i, j^\prime k^\prime} (x, y)& =& \frac{\,R^4}{6(m+4)(m+\frac{15}2)}
\eta _{j^{\prime }k^{\prime }}(\partial _\lambda +\frac 1\lambda )(\frac{\partial _{
\lambda ^{\prime }}}{\lambda ^{\prime }}-\frac{3+m}{\lambda ^{2\prime }})\partial _i
\Delta _{\nu _0}(p)\nonumber \\
&& +\frac{\,R^4}{6(m+4)(m+\frac{15}2)}\partial _i\partial _{j^{\prime }}
\partial _{k^{\prime }}(\partial _\lambda +\frac 1\lambda )\Delta _{\nu _0}(p)\nonumber \\
&&- \frac{2R^4}{3(m+6)(m+4)}\eta _{j^{\prime }k^{\prime }}\partial _i(\partial _\lambda 
-\frac 2\lambda )(\partial _{\lambda ^{\prime }}-\frac m{2\lambda ^{2\prime }})\Delta _
\nu (p)\nonumber\\
&& - \frac{R^4}{2\,(m+6)}\eta _{i\,j^{\prime }}\partial _{k^{\prime }}\frac 1{\lambda 
\,\lambda ^{\prime }}(\partial _{\lambda ^{\prime }}-\frac 3{\lambda ^{\prime }})\Delta 
_\nu (p)\nonumber\\
&&-\frac{R^4}{2\,(m+6)}\eta _{i\,k^{\prime }}\partial _{j^{\prime }}\frac 1{
\lambda \,\lambda ^{\prime }}(\partial _{\lambda ^{\prime }}-\frac 3{\lambda ^{\prime }})
\Delta _\nu (p)\nonumber\\
&&-\frac{2R^4}{3(m+6)(m+4)}(-\frac 32+\frac 3{4\nu }(m+3))\partial _i
\partial _{j^{\prime }}\partial _{k^{\prime }}\frac 1{\,\lambda ^{\prime }}\Delta 
_{\nu +1}(p)\nonumber\\
&&-\frac{2R^4}{3(m+6)(m+4)}(-\frac 32-\frac 3{4\nu }(m+3))\partial _i\partial _
{j^{\prime }}\partial _{k^{\prime }}\frac 1{\,\lambda ^{\prime }}\Delta _{\nu -1}(p)
\nonumber \\
&&+\frac{2R^4}{3(m+6)(m+4)}\partial _i\partial _{j^{\prime }}\partial _{k^{\prime }}
(\partial _\lambda -\frac 2\lambda )\Delta _\nu (p) \label{61}\\
\Delta_{i j, k^\prime l^\prime} (x, y) &= &
\eta _{i\,j}\eta _{k^{\prime \,}l^{\prime }}\left\{ \frac{\,R^4}{6(m+4)(m+\frac{15}2)}(\lambda \partial _\lambda -3-m)(\lambda ^{\prime }\partial _{\lambda ^{\prime }}-3-m)\Delta _{\nu _0}(p)\right.\nonumber\\
&&\qquad \qquad \left.+\frac{2R^4}{3(m+6)(m+4)}(\frac{\partial _\lambda }\lambda +\frac m{2\lambda ^2})(\frac{\partial _{\lambda ^{\prime }}}{\lambda ^{\prime }}+\frac m{2\lambda ^{2\prime }})\Delta _\nu (p)\right.\nonumber\\
&&\qquad \qquad \left.-\frac{R^4}{2(\lambda \lambda ^{\prime })^2}\Delta _\nu (p)\right\}\nonumber\\ 
&&+\eta _{i\,j}\partial _{k^{\prime }}\partial _{l^{\prime }}\left\{ \frac{\,R^4}{6(m+4)(m+\frac{15}2)}\frac{(\lambda \partial _\lambda -3-m)}{\lambda ^2}\Delta _{\nu _0}(p)\right.\nonumber\\
&&\qquad \qquad \left. +\frac{2R^4}{3(m+6)(m+4)}(\frac{\partial _\lambda }\lambda +\frac m{2\lambda ^2})\Delta _\nu (p)\right.\nonumber\\
&&\qquad \qquad \left. -\frac 1{\lambda \lambda ^{\prime }}\frac{2R^4}{3(m+6)(m+4)}\frac{24m+90-36\nu -12m\nu }{16\nu ^2}\Delta _{\nu -1}(p)\right.\nonumber\\
&&\qquad \qquad\left. -\frac 1{\lambda \lambda ^{\prime }}\frac{2R^4}{3(m+6)(m+4)}\frac{24m+90+36\nu +12m\nu }{16\nu ^2}\Delta _{\nu +1}(p)\right\}\nonumber\\
&&+\eta _{k^{\prime }l^{\prime }}\partial _i\partial _j\left\{ \frac{\,R^4}{6(m+4)(m+\frac{15}2)}\frac{(\lambda ^{\prime }\partial _{\lambda ^{\prime }}-3-m)}{\lambda ^{2\,\prime }}\Delta _{\nu _0}(p)\right. \nonumber\\
&&\qquad \qquad \left.+\frac{2R^4}{3(m+6)(m+4)}(\frac{\partial _{\lambda ^{\prime }}}{\lambda ^{\prime }}+\frac m{2\lambda ^{2^{\prime }}})\Delta _\nu (p)\right.\nonumber\\
&&\qquad \qquad \left.-\frac 1{\lambda \lambda ^{\prime }}\frac{2R^4}{3(m+6)(m+4)}\frac{24m+90-36\nu -12m\nu }{16\nu ^2}\Delta _{\nu -1}(p)\right.\nonumber\\
&&\qquad \qquad \left.-\frac 1{\lambda \lambda ^{\prime }}\frac{2R^4}{3(m+6)(m+4)}\frac{24m+90+36\nu +12m\nu }{16\nu ^2}\Delta _{\nu +1}(p)\right\}\nonumber\\ 
&&-(\eta _{il^{\prime }}\partial _j\partial _{k^{\prime }}+\eta _{ik^{\prime }}\partial _j\partial _{l^{\prime }}+\eta _{jl^{\prime }}\partial _i\partial _{k^{\prime }}+\eta _{jl^{\prime }}\partial _i\partial _{k^{\prime }})\nonumber\\
&&\qquad \qquad \frac{R^4}{2\,(m+6)}\frac 1{\lambda \lambda ^{\prime }} ( \frac{12\nu +8m+30}{16\nu ^2}\Delta _{\nu -1}(p)
+\frac{-12\nu +8m+30}{16\nu ^2}\Delta _{\nu +1}(p) ) \nonumber\\
&&+(\eta _{il^{\prime }}\eta _{jk^{\prime \,}}+\eta _{ik^{\prime }}\eta _{jl^{\prime \,}})\frac{R^4}{2(\lambda \lambda ^{\prime )^2}}\Delta _\nu (p)\nonumber\\
&&+\partial _i\partial _j\partial _{k^{\prime }}\partial _{l^{\prime }}\frac{2R^4}{3(m+6)(m+4)} \quad *\nonumber\\
&&\left\{ \frac{m+6}{4(m+\frac{15}2)}\Delta _{\nu _0}(p)+\frac{4+m}{19+4m}\Delta _\nu (p)\right.\nonumber\\
&&\left.+\frac 1{4(\nu -1)^2}\left\{ \frac{-3(4+m)(6+m)}{60+16m}-3(m+4)(\frac 12+\frac 3{4\nu })^2-\frac{9(m+3+\nu )}{60+16m}\right\} \Delta _{\nu -2}(p)\right.\nonumber\\
&&\left.+\frac 1{4(\nu -1)^2}\left\{ \frac{3(4+m)(6+m)}{60+16m}-3(m+4)(-\frac 12+\frac 3{4\nu })^2-\frac{9(m+3-\nu )}{60+16m}\right\} \Delta _{\nu +2}(p)\right\}\nonumber\\
\label{62}
\end{eqnarray}
\normalsize
\bigskip

\noindent where 
\begin{eqnarray}
 \Delta _{\nu_{0}} (p)&=&i\frac \pi {4R^2}\frac{(\lambda \lambda ^ {\prime} )^{3/2}}
{(2\pi )^3}\int d^3ke^{i\vec k\cdot (\vec x-\vec y)}({\cal H}_{\nu_{0}} ^{(1)}(\lambda k)
{\cal H}_{\nu_{0}} ^{(2)}(\lambda ^{\prime} k)-{\cal H}_{\nu_{0}} ^{(2)}(\lambda k){\cal H}_{\nu_{0}} ^{(1)}(\lambda 
^{\prime }k))\nonumber\\
&=&\frac{-1}{8\pi R^2}\frac{(\frac 14-\nu_{0} ^2)}{\cos (\nu_{0} \pi )}
\epsilon (\lambda -\lambda ^{\prime} )\Im \, \left[_2F_1(\frac 32-\nu_{0} ,\frac 32+\nu_{0} ,2,\frac{1+p}2)\right]  \label{63}
\end{eqnarray}
and a similar representation for 
$\Delta_{\mu \nu, \rho \sigma}^1 (x, y)$ and $\Delta_{\mu \nu, \rho \sigma}^F (x, y)$ where
$\Delta_{\nu_{0}}$ is replaced respectively by:
\begin{eqnarray}
\Delta _{\nu_{0}}^{(1)} (p)&=&\frac \pi {4R^2}\frac{(\lambda \lambda ^{\prime} )^{3/2}}
{(2\pi )^3}\int d^3ke^{i\vec k\cdot (\vec x-\vec y)}({\cal H}_{\nu_{0}} ^{(1)}(\lambda k)
{\cal H}_{\nu_{0}} ^{(2)}(\lambda ^{\prime} k)+{\cal H}_{\nu_{0}} ^{(2)}(\lambda k){\cal H}_{\nu_{0}} ^{(1)}(\lambda 
^{\prime }k))\nonumber\\
&=&\frac{1}{8\pi R^2}\frac{(\frac 14-\nu_{0} ^2)}{\cos (\nu_{0} \pi )}
\Re \, \left[_2F_1(\frac 32-\nu_{0} ,\frac 32+\nu_{0} ,2,\frac{1+p}2)\right]  \label{63b}
\end{eqnarray}
and
\begin{eqnarray}
\Delta ^F _{\nu_{0}} (p)=\frac 1{16\pi R^2}\frac{(\frac 14-\nu_{0} ^2)}{\cos (\nu_{0} \pi )}\,_2F_1(\frac 32-\nu_{0} ,\frac 32+\nu_{0} ,2,\frac{1+p}2-i\epsilon) \label{plus}
\end{eqnarray}
The occurrence of the factor $\mbox{\rm{sec}}(\nu \pi)$ in eqs(\ref{63},\ref{63b}) implies that the special values of 
$\nu = n + \frac 1 2 \quad ,  \mbox{\rm{i.e.}} \quad {m+4=n(n+1)}$ need a
special analysis. These values of the index of the modes correspond to eigenvalues of the
Laplace-Beltrami operator on the 4-sphere $S^4$, i.e. situations where the analytic 
continuation of the propagators built on $S^4$ onto de Sitter space fails \cite{AnMo}. In these cases it doesn't exist 
de Sitter invariant states; the invariant propagators describe expectations values of the field with respect to a density matrix.  
\section{Invariant representation of the Green functions}
 The $ (\lambda, \vec x)$ coordinates (\ref{13}) cover only one half of de
Sitter space corresponding to the causal past of a physical observer
(region ${\cal{O}}$ on fig.1). This domain is bounded by a future event horizon
($X^0=X^4$ in eq.(\ref{13})) and is invariant under a seven parameter subgroup of the full de Sitter
group  O (4, 1). This group is isomorphic to $ {\bf {R}}_0^+ \times E(3)$, as it is obvious
from the writing (\ref{14}) of the metric. It consists of the $E(3)$ euclidean motions
preserving $\sum_i (dx^i)^2$ in eq.(\ref{14}) and the dilatations $ \lambda \mapsto k \lambda, \vec x
\mapsto k \vec x$.

\noindent On this domain we may express the Green functions in terms of obviously
geometrically invariant quantities. Let us consider two points ($ P, Q$) belonging
to ${\cal{O}}$ and denote by $X^A$ and $ Y^A \quad ( A = 0,...,4)$ their coordinates in the
embedding $ M^5$ space. The tangent vectors at the ends of the unique geodesic
in ${\cal{O}}$ joining them are:

\begin{eqnarray}
T_P^A &=& {Y_A - p X_A \over R \vert p^{2} -1 \vert^{1/2}} \; , \nonumber \\
T_Q^A &=& {- X^A + p Y^A \over R \vert p^{2} -1 \vert^{1/2}}  \quad ,
 \label{67}
\end{eqnarray}
where:
\begin{eqnarray}
p = {\eta_{AB} X^A Y^B \over R^2} \equiv {X.Y \over R^2} = \frac{\lambda_P ^2+\lambda_Q^2-(\vec x_P-\vec x_Q)^2}{2\lambda_P \lambda_Q}. \label{p}
\end{eqnarray}
\bigskip

\noindent The components $ V^A_Q$ of the vector $ V^A_P$ parallely transported
from $ P \;{\rm to}\; Q $ are

\begin{eqnarray}
V_Q ^A = V_P^A - {(T_P.V) \over T_P. T_P} \; T_P + {V.T_Q \over T_Q. T_Q}
\; T_Q  \label{68}
\end{eqnarray}

\bigskip

\noindent and $ V_Q ^A = V_P^A$ when the geodesic is a null one $( T_P.T_P = 0
= T_Q.T_Q)$

\bigskip

\noindent We deduce immediately from this the expression the $ M^5$
components of the tensor of parallel transport from $P$ to $Q$ :

\begin{eqnarray}
\Theta_B^{A^\prime} = \delta_B^{A^\prime} - {X^{A^\prime} X_B + Y^{A^\prime}
Y_B + X^{A^\prime} X_B - p Y^{A^\prime} Y_B \over R^2 (p + 1) } \; .
\label{69}
\end{eqnarray}

\bigskip

\noindent In $ (\lambda, \vec x)$ coordinate, with $ \vec r = \vec x_Q -
\vec x_P$, the components of these objects read:

\begin{eqnarray}
T_P^\alpha& = &\left( {\lambda_Q^2 - \lambda_P^2 - r^2 \over 2 \lambda_Q}\; ,
{\lambda_P \over \lambda_Q} \; r^i \right) {1 \over R^2 \mid p^2 -
1 \mid^{1/2}} \nonumber \\
T_Q^{\alpha^\prime}& =& \left( - {\lambda^2_P - \lambda_Q^2 - r^2 \over 2
\lambda_P} \; , {\lambda_Q \over \lambda_P} r^i\right) \; {1 \over R^2 \mid p^2 -
1\mid ^{1/2}} 
\label{70}
\end{eqnarray}

\noindent and
\bigskip
\begin{eqnarray}
\Theta_\beta^{\alpha^\prime} = \left(\begin{array}{cc}{\lambda_Q \over
\lambda_P} \; {(\lambda_Q + \lambda_P)^2 + r^2 \over (\lambda_Q + \lambda_P)^2 -
r^2} & r^i\; {(\lambda_Q + \lambda_P) \over \lambda_P^2 (p + 1)} 
\\
{r ^{j^\prime} (\lambda_Q + \lambda_P) \over \lambda_P^2 (p+1)} & {\lambda_Q
\over \lambda_P} \delta^{i_{j^\prime}} + {r^i r^{ j^\prime} \over \lambda_P^2
(p+1)}\\ 
\end{array}\right)\; .
\label{71}
\end{eqnarray}

\noindent Following Allen \cite{Allen2} we have to consider the five invariant bitensors 
defined by:
\begin{eqnarray}
O_1^{\alpha \beta, \mu^\prime \nu^\prime} &=& g^{\alpha \beta} g^{\mu^ \prime \nu^\prime}\qquad,
\nonumber \\
O_2^{\alpha \beta, \mu^\prime \nu^\prime} &=& T_P^\alpha T_P^\beta T_Q^{\mu^\prime}
T_Q^{\nu^\prime} \qquad, \nonumber \\
O_3^{\alpha \beta, \mu^\prime \nu^\prime} &=&( \Theta^{\alpha \mu^\prime} \Theta^{\beta
\nu^\prime} + \Theta^{\alpha \mu^\prime} \Theta^{\beta \nu^\prime}) \qquad, \nonumber \\
O_4 ^{\alpha \beta, \mu^\prime \nu^\prime}&=& ( g^{\alpha \beta}\; T_Q^{\mu^\prime}\;
T_Q^{\nu^\prime} \;T_P^\alpha \;T_P^\beta g^{\mu^\prime \nu^\prime}) \qquad ,\nonumber \\
O_5^{\alpha \beta, \mu^\prime \nu^\prime} &=& 4 T_P^{(\alpha} \Theta^{\beta)(\mu^\prime}
T_Q^{\nu^\prime)} \qquad.
\label {72}
\end{eqnarray}

\bigskip

\noindent Using the previous expressions of the components of $T_P, T_Q$ and $ \Theta$
and expliciting the action of the derivatives in eqs (\ref{57}) to (\ref{62}) we obtain by identification
invariant expressions of the Green functions. The details of the calculations
are very tedious and we don't reproduce them here. To illustrate the method, we
shall consider only the case of the massive vector field. It has been demonstrated
in \cite{SpTh} that the $(\lambda,\lambda ^{\prime})$  component (in the coordinates
system (\ref{13}) ) of the Feynman
propagator, defined by the equation $(g_{\alpha \beta}\Box-R_{\alpha \beta}-M^2 g_{\alpha \beta})
\Delta^{F \, \beta \gamma ^{\prime}}=0$ is given by:
\begin{eqnarray} 
{\Delta^F}^{\lambda}_ {\lambda ^{ \prime} }(x,y )=-\frac {R^2}{\lambda ^{\prime 2}} \frac 1{M^2}(\vec \nabla _x\cdot \vec \nabla _y)\left[ (\frac{\lambda \lambda ^{\prime} }{R^2})^2 \Delta ^F _\sigma (p)\right]\label{truc}\qquad. 
\end{eqnarray}
with:
\begin{eqnarray} 
\sigma=i \;\sqrt{M^2 R^2 - {1 \over 4}} \quad .
\end{eqnarray}
\noindent The others components are given by similar expressions (differential operators
acting on invariant functions of $p$, see \cite{SpTh}). On the other hand, the most general maximally
symmetric invariant bitensor with the same index structure as the vectorial propagator reads as the
combination:
\begin{eqnarray}
\Theta^{\alpha}_ {\alpha \prime}F(p)+T_P^\alpha {T_Q}_{\alpha \prime} G(p) 
\end{eqnarray}
So, the vectorial invariant Feynman propagator ${\Delta^F}^{\beta}_{ \gamma \prime}$ can be written as:
\begin{eqnarray}
{\Delta^F}^{\beta}_{ \gamma \prime}=\Theta^{\beta}_{ \gamma \prime}\alpha (p)+T_P^\beta {T_Q}_{\gamma \prime} \beta (p)\label{combi}
\end{eqnarray}
\noindent for some functions $\alpha(p), \beta(p)$.
\bigskip
\noindent Equation (\ref{truc}) can be reexpressed as a function of $p$ and $\xi={\lambda \over \lambda ^{\prime}}$:
\begin{eqnarray} 
{\Delta^F} ^{\lambda}_{ \lambda ^{\prime }}(x,y)&=&- \frac 1{M^2}(\frac{\lambda}{R }^2\left[ \frac 3{\lambda \lambda ^{\prime }}\frac d{dp}-(\frac r{\lambda \lambda ^{\prime }})^2\frac{d^2}{dp^2}\right] \Delta ^F _\sigma (p)\nonumber\\
&&= -\frac 1{R^2 M^2}\left[ 3\xi \frac d{dp}-\xi(2p-\xi-\xi^{-1})\frac{d^2}{dp^2}\right] \Delta ^F _\sigma (p)
\end{eqnarray}
thanks to eq.(\ref{p}). 
\noindent Comparing this expression with the $(\lambda,\lambda^{\prime})$  component of
eq.(\ref{combi}), in which the terms are grouped together according to their powers of $\xi$,
we may identify the coefficients $\alpha (p)$ and $\beta (p)$ of the
decomposition:   
\begin{eqnarray}
\alpha (p)&=&\frac 1{m^2R^2}\left[ 3p\frac d{dp}+(p^2-1)\frac{d^2}{dp^2}\right] \Delta ^F _\sigma (p) \quad,\\
\beta (p)&=&\frac 1{m^2R^2}\left[ 3(1-p)\frac d{dp}+(1-p^2)\frac{d^2}{dp^2}\right] \Delta ^F _\sigma (p) \quad,
\end{eqnarray}
\noindent and possibly use the other components of the propagator to check
the results. They are in agreement with those given in \cite{Allen2}. \\
\\
\bigskip
\noindent A similar calculation involving only two types of components:
$\Delta^{00,0 ^{\prime} i^{\prime}}\; \mbox{\rm{and}}\; \Delta^{00,i ^{\prime},j^{\prime}}$
allows to determine the invariant form of the propagator:
\begin{eqnarray}
\Delta^{F\, \mu \nu ,\rho ^{\prime }\sigma ^{\prime }}(p)=\alpha (p)O_1^{\mu \nu ,\rho ^{\prime }\sigma ^{\prime }}+\beta (p)O_2^{\mu \nu ,\rho ^{\prime }\sigma ^{\prime }}+\gamma (p)O_3^{\mu \nu ,\rho ^{\prime }\sigma ^{\prime }}+\delta (p)O_4^{\mu \nu ,\rho ^{\prime }\sigma ^{\prime }}+\epsilon (p)O_5^{\mu \nu ,\rho ^{\prime }\sigma ^{\prime }}  
\end{eqnarray}
\noindent with:
{\footnotesize
\\
\\
\begin{eqnarray}
\alpha (p) &=&\frac 2{3(m+4)(m+6)}\left\{ \frac{-(m+6)\left[ (p^2-1)(m+8)+2\right] \Delta _\nu ^F-p\left[ 2(m+6)(p^2-1)+8\right]\Delta _\nu ^{F\prime}}{2(p^2-1)}\right.\nonumber\\
&&\qquad \qquad \qquad \qquad \left. + \frac{\left[ (p^2-1)(m^2+5m+9)-m\right] \Delta _{\nu _0}^F+\left[ p(p^2-1)(2m+3)-4p\right] \Delta _{\nu _0}^{F\prime}}{p^2-1}\right\}\\
\beta (p) &=&\frac 2{3(m+4)(m+6)} \cdot \nonumber \\
 &&\left\{ \frac{-(m+6)\left[ (m+8)(p^2-1)-20p-28\right] \Delta _\nu ^F+\left[ (p^2-1)(2p(m+6)+10(m+2))-112p-80\right] \Delta _\nu ^{F\prime}}{p^2-1}\right.\nonumber\\
&&\left. +\frac{\left[ (p^2-1)(m^2-16m)-2m(10p+14)\right] \Delta _{\nu _0}^F+\left[ (p^2-1)(8mp-48p+4m-64)-80-112p\right] \Delta _{\nu _0}^{F\prime}}{p^2-1}\right\} \nonumber\\
\\ 
\gamma (p) &=&\frac 2{3(m+4)(m+6)}\left\{ \frac{(m+6)\left[ 3(p^2-1)(m+8)-4\right] \Delta _\nu ^F+\left[ (p^2-1)6p(m+6)-16p\right] \Delta _\nu ^{F\prime}}{4(p^2-1)}\right.\nonumber\\
&&\qquad \qquad \qquad \qquad \left.-\frac{m\Delta _{\nu _0}^F+4p \Delta _{\nu _0}^{F\prime}}{p^2-1}\right\}\\ 
\delta (p) &=&\frac{2\,\varepsilon(p^2-1)}{3(m+4)(m+6)}\left\{ \frac{-(m+6)\left[ (p^2-1)(m+8)+12\right] \Delta _\nu ^F-\left[ (p^2-1)(m+6)2p+48p\right] \Delta _\nu ^{F\prime}}{2(p^2-1)}\right.\nonumber\\
&&\qquad \qquad \qquad \qquad \left.+\frac{m\left[ (p^2-1)(m-1)-6\right] \Delta _{\nu _0}^F+p\left[ (p^2-1)5m-24\right] \Delta _{\nu _0}^{F\prime}}{p^2-1}\right\}\\ 
\epsilon (p) &=&\frac{2\,\varepsilon(p^2-1)}{3(m+4)(m+6)}\cdot \nonumber\\
&&\left\{ \frac{(m+6)\left[ 3(p^2-1)(m+8)-20p-4\right] \Delta _\nu ^F+\left[ (p^2-1)(6p(m+6)+10(m+2))-16(p+5)\right] \Delta _\nu ^{F\prime}}{4(p^2-1)}\right.\nonumber\\
&&\left.+\frac{-m(1+5p)\Delta _{\nu _0}^F+\left[ (p^2-1)(m-16)-4(p+5)\right] \Delta _{\nu _0}^{F\prime}}{p^2-1}\right\}\nonumber\\ 
\end{eqnarray}}
where $\Delta_\nu ^{F\prime}=\frac d{dp}\Delta_\nu ^F$
\pagebreak
\section{Scalar Green function revisited}
\noindent In ref \cite{ScSp1} we have shown that the Green's functions of the scalar field
equation on the domain ${\cal{ O}}$~:

\begin{eqnarray}
( \Box - M^2) \varphi \equiv \left( {\lambda^2 \over R^2} ( - \partial_\lambda^2 + \vec
\nabla^2) + {2 \lambda \over R} \partial_\lambda - M^2 \right) \varphi = 0
\end{eqnarray}

\bigskip

\noindent can be written as a superpositions of modes expressed in terms of Hankel 
functions~:

\begin{eqnarray}
u_{\vec p} (\lambda) &=& {\surd {\pi} \over {2R}} \lambda^{3/2} \left[ c (\vec p) {\cal H}
_{\nu_0} ^{(1)} ( \lambda p) + d ( \vec p) {\cal H} _{\nu_0} ^{(2)} ( \lambda p) \right] e^{i \vec
p. \vec x} \nonumber \\
\nu_0&=& i \sqrt{m^2 R^2 - {9 \over 4}}  
\end{eqnarray}

\bigskip

\noindent with $ \vert d (\vec p) \vert^2 - \vert c (\vec p) \vert^2 = 1,\quad \mbox{\rm {and  }}  c (\vec
p) d (- \vec p) - c (- \vec p) d (\vec p) = 0 \; ,$ the last conditions resulting from the normalisation condition $ u_{\vec p} *
u_{\vec p^\prime} = \delta^3 (\vec p - \vec p^\prime) \; .$ These conditions are not sufficient to fix the vacuum (the positive frequency modes). If
we impose the vacuum to be invariant with respect to the 7-parameter isometry group of
${\cal{ O}}$, extra (necessary) conditions appear. The coefficients $ c(
\vec p)$ and $ d (\vec p)$ have to be constant. The resulting Green functions
still depend on three parameters. Definite values of these parameters are obtained by
imposing that the short distance singularities of the Feynman propagator are the same as
in flat space:
\begin{eqnarray}
\lim \limits_{\sigma \rightarrow 1}\sigma ^2\Delta ^F(\sigma )=\frac{-1}{2\pi ^2}, \quad \mbox{\rm where} \quad p=\mbox{\rm{cosh}}({{\sigma} \over R})
\end{eqnarray}
Then one obtains:
\begin{eqnarray}
c = 0\; , \quad d = 1 \quad ,
\end{eqnarray}

\bigskip

\noindent because the phase of $d$ becomes irrelevant. The Feynman propagator for $M \neq 0$, is
given by:

\begin{eqnarray}
\Delta^{F} (x, y) = {1 \over 16 \pi R^2} \; {(M^2 R^2 - 2) \over \cos \nu_0 \pi} \; F
\left( {3 \over 2} + \nu_0, {3 \over 2} - \nu_0; 2; {{1 + p \over 2} - i \epsilon } \right) \; .\label{76} 
\end{eqnarray}

\bigskip

\noindent while for $ M = 0$ one obtains \cite{ScSp1} :

\begin{eqnarray}
\Delta^{F} (x, y) &=& {1 \over 4 \pi^2 R^2} \left[ {1 \over 1 - p} - ln \left| {\lambda
\lambda^\prime \over R^2} (p - 1) \right| + c^{te} \right] \nonumber \\
& & - {i \over 4 \pi R^2} \varepsilon (\lambda - \lambda^\prime) (\delta (p - 1) + \theta (p
- 1))
\end{eqnarray}

\bigskip

\noindent which is only $ E(3)$ invariant. We plan to discuss the physical significance
of this choice of modes in a forthcoming publication \cite{GaSp}.

Up to now, all the expressions of the modes that we have considered where defined only on 
${\cal{O}}$. If we extend the definition of $ \lambda$ and $ \vec x$ by eq.(\ref{13}) on the full de Sitter
space (excepted on the horizon $H$, i.e. the 3-surface $X_4=X_0$) we may analytically 
continue the modes by considering the behaviour of a wave packet near $H$.
Typically, such wave packet behaves as:
\begin{eqnarray}
\varphi(\lambda ,\vec x) &=& \int \frac{\sqrt{\pi }}{2R}\left| 
\lambda \right| ^{\frac 32}{\cal{H}}_\nu ^2(\lambda p)
\frac{e^{i\vec p\cdot \vec x}}{(2 \pi )^{\frac 32}} f\,(\vec p)d^3p 
\nonumber\\
&\sim& \int \frac{\left| \lambda \right| }{\sqrt{2\pi }R}e^{i\frac \pi 4}e^{-i(p\lambda -\vec p\cdot \vec x)}f\,(\vec p)d^3p \label{wp}
\end{eqnarray}
\noindent The Kirchoff (Poisson) formula giving the solution of the massless scalar wave equation in flat
space insures that this expression remains finite on the horizon, despite the presence of the
divergent factor $\left| \lambda \right| $.
\noindent The continuation of the modes across the horizon is obtained by looking in the region
$\lambda<0$ which combination of Hankel functions have an asymptotic expansion that matches with the
one used in eq.(\ref{wp}). This leads us immediately to an expression of the modes valid on the
full de Sitter space:
\begin{eqnarray}
u_{\vec p} (\lambda) = {\sqrt{\pi} \over 2 R} \vert \lambda \vert^{3/2} [ \theta
(\lambda) {\cal H}_{v_0}^{(2)} (( \lambda-i \epsilon) p)  - i \theta (-\lambda) {\cal
H}_{v_0}^{(1)} ( -( \lambda-i\epsilon) p)] \frac {e^{i \vec p. \vec x}}{(2\pi)^{\frac 3 2}}  \; . 
\end{eqnarray}

\noindent Inserting this expressions of modes in integrals like those considered in eqs (\ref{63}, \ref{63b}), 
we conclude that the expression (\ref{76}) is valif on whole de Sitter space, with $p$
still given by eq.(\ref{p}) whatever are the signs of $\lambda$ and $\lambda ^\prime$

Note that the region $\bar {\cal{O}}=H^4\setminus{\cal{O}}$ is isometric to ${\cal{O}}$ but
with time running in the opposite way. This is in accord with the fact that it is precisely the
Hankel function $ {\cal H}^{(1)} (\vert \lambda p\vert )$ which is coupled to $ {\cal H}^{(2)} (\vert \lambda
p \vert)$, these two functions being of opposite frequences on both ${\cal{O}}$ and $\bar {\cal{O}}$.
Finally, note also that by the continuation $ R \mapsto i \, R$ we obtain invariant
expressions of Green functions on anti-de Sitter space.
\bigskip
\section*{Acknowledgments}
\noindent We would like to thank R.Brout, C. De Mol, R. Parentani and S. Massar for helpful discussions
during the elaboration of this paper. One of us (Cl.G.) would like to thank the
Universit\'e de Mons-Hainaut and the Fonds National de la Recherche Scientifique for financial support.
\pagebreak
\section*{Appendix}
\renewcommand\theequation{A.\arabic{equation}}
\setcounter{equation}{0}
\noindent We have collected in this appendix the explicit expressions of the modes we have used to obtain the propagators. They are expressed for each $ \vec p$ in the
base $\left\{ \vec e_0=\vec \partial _0,\vec e_1,\vec e_2,
\vec e_3=\frac{\vec p}p\right\} $.
They read as:
\footnotesize{
\begin{eqnarray}
\hbar_{\mu \nu }^S(\vec p,\lambda ,\vec x)&=&
\left(\begin{array}{llll} O_1 & 0 & 0 & O_3
 \\ 0 & O_7 & 0 & 0 \\ 0 & 0 & O_7 & 0 \\ O_3 & 0 & 0 & O_5\end{array} \right)
\frac 1{(\lambda p)^2}\frac{\sqrt{p}}{N_S} e^{i\vec p\cdot \vec x}(\lambda p)^{\frac 32}{\cal{H}}_{i\sqrt{m-\frac 94}}^{(2)}(\lambda p)\\
\hbar_{\mu \nu }^{TT}(\vec p,\lambda ,\vec x)&=&
\left(\begin{array}{llll} 1 & 0 & 0 & O_2 
 \\ 0 & O_6 & 0 & 0 \\ 0 & 0 & O_6 & 0 \\ O_2 & 0 & 0 & O_4\end{array} \right)
\frac 1{(\lambda p)^2}\frac{\sqrt{p}}{N_{TT}}e^{i\vec p\cdot \vec x}(\lambda p)^{\frac 72}{\cal{H}}_{i\sqrt{m+\frac {15}4}}^{(2)}(\lambda p)\\
\hbar_{\mu \nu }^{\perp 1}(\vec p,\lambda ,\vec x)&=&
\left(\begin{array}{llll} 0 & 1 & 0 & 0 
 \\ 1 & 0 & 0 & O_2 \\ 0 & 0 & 0 & 0 \\ 0 & O_2 & 0 & 0 \end{array} \right)
\frac 1{(\lambda p)^2}\frac{\sqrt{p}}{N_{\perp}} e^{i\vec p\cdot \vec x}(\lambda p)^{\frac 52}{\cal{H}}_{i\sqrt{m+\frac {15}4}}^{(2)}(\lambda p)\\
\hbar_{\mu \nu }^{\perp 2}(\vec p,\lambda ,\vec x)&=&
\left(\begin{array}{llll} 0 & 0 & 1 & 0 
\\ 0 & 0 & 0 & 0 \\ 1 & 0 & 0 & O_2 \\ 0 & 0 & O_2 & 0\end{array} \right) 
\frac 1{(\lambda p)^2}\frac{\sqrt{p}}{N_{\perp}} e^{i\vec p\cdot \vec x}(\lambda p)^{\frac 52}{\cal{H}}_{i\sqrt{m+\frac {15}4}}^{(2)}(\lambda p)\\
\hbar_{\mu \nu }^{\perp \perp 1}(\vec p,\lambda ,\vec x)&=&
\left(\begin{array}{llll} 0 & 0 & 0 & 0 
\\ 0 & 1 & 0 & 0 \\ 0 & 0 & -1 & 0 \\ 0 & 0 & 0 & 0\end{array} \right)
\frac 1{(\lambda p)^2}\frac{\sqrt{p}}{N_{\perp \perp}} e^{i\vec p\cdot \vec x}(\lambda p)^{\frac 32}{\cal{H}}_{i\sqrt{m+\frac {15}4}}^{(2)}(\lambda p)\\
\hbar_{\mu \nu }^{\perp \perp 2}(\vec p,\lambda ,\vec x)&=&
\left(\begin{array}{llll} 0 & 0 & 0 & 0
 \\ 0 & 0 & 1 & 0 \\ 0 & 1 & 0 & 0 \\ 0 & 0 & 0 & 0\end{array} \right)
\frac 1{(\lambda p)^2}\frac{\sqrt{p}}{N_{\perp \perp}}e^{i\vec p\cdot \vec x}(\lambda p)^{\frac 32}{\cal{H}}_{i\sqrt{m+\frac {15}4}}^{(2)}(\lambda p)
\end{eqnarray}}
\normalsize
\noindent where:
\begin{eqnarray}
O_1&=&(3\lambda \partial _\lambda -(\lambda p)^2+3)\\
O_2&=&\frac{-i}p(\partial _\lambda -\frac 4\lambda )\\
O_3&=&ip\lambda ^2(\partial _\lambda +\frac 1\lambda )\\
O_4&=&(\frac 2{\lambda p^2}\partial _\lambda +1+\frac{m-4}{(\lambda p)^2})\\
O_5&=&(\lambda \partial _\lambda -((p\lambda )^2+m+3))\\
O_6&=&(-\frac 1{p^2\lambda }\partial _\lambda +\frac{4-m}{2(p\lambda )^2})\\
O_7&=&(\lambda \partial _\lambda -3-m)
\end{eqnarray}
These modes are orthogonal for the scalar product (\ref{9}). In order that they satisfy the orthonormalization
condition (\ref{10}), their coefficients must be chosen, up to a phase, as:
\begin{eqnarray}
N_S&=&\frac 1R\sqrt{208\pi ^2(m+4)(m+\frac{15}2)} \\
N_{TT}&=&\frac 1R\sqrt{48\pi ^2(m+6)(m+4)}\\
N_{\perp}&=&\frac 1R\sqrt{64\pi ^2(m+6)}\\
N_{\perp \perp}&=&\frac 1R\sqrt{64\pi ^2}
\end{eqnarray}

\pagebreak

\pagebreak
\section*{Figure Caption}
Penrose diagram of de Sitter space \cite{HaEl}. Region ${\cal{O}}$  ($\lambda > 0$) corresponds to
the causal past of observers $\vec x = cst, \quad \lambda > 0$. Their common future event horizon
${\cal{H}}$ is the boundary of the two coordinate patches $(\lambda>0, \vec x) \quad \mbox{\rm{and}} \quad 
(\lambda<0, \vec x)$. Dashed lines represent $\vec x = cst$ world lines.
\end{document}